\documentclass[letter]{aa}
\usepackage{graphicx}
\usepackage{txfonts}
\usepackage{lipsum}
\usepackage{subcaption}
\usepackage{lscape}             
\usepackage{placeins}           
                                
\usepackage{xcolor}
\usepackage[normalem]{ulem}

\usepackage{etoolbox}
\newcommand{\sugg}[3]{
  {\ifthenelse{\equal{#2}{}}%
   {\textcolor{#1}{#3}}%
   {\textcolor{#1}{\sout{#2} #3}}%
  }
}

\newcommand{\tcore}{$\tilde{T}_{\mathrm{core}}$}

\newcommand{\nscool}{\texttt{NSCool}}

\newcommand{\mxb}{MXB 1659-29}
\newcommand{\mxbi}{MXB 1659 (OI)}
\newcommand{\mxbii}{MXB 1659 (OII)}
\newcommand{\mxbbrief}{MXB 1659}
\newcommand{\ks}{KS 1731-260}
\newcommand{\ksbrief}{KS 1731}
\newcommand{\exo}{EXO 0748-676}
\newcommand{\exobrief}{EXO 0748}
\newcommand{\xte}{XTE J1701-462}
\newcommand{\xtebrief}{XTE J1701}
\newcommand{\rxs}{1RXS J180408.9-342058}
\newcommand{\rxsbrief}{1RXS J1804}
\newcommand{\igr}{IGR J17480-2446 (Ter5 X-2)}
\newcommand{\igrbrief}{Ter5 X-2}
\newcommand{\swift}{Swift J174805.3-244637 (Ter5 X-3)}
\newcommand{\swiftbrief}{Ter5 X-3}
\newcommand{\tbteff}{$T_{b} - T_{\text{eff}}$}
\newcommand{\tblbmdot}{$T_{b} - L_{b} - \dot{M}$}

\newcommand{\twm}{watershed model}

\newcommand{\fonto}[1]{#1}

\begin{document}

   \title{On a re-examination of neutron star cooling in transient sources}

   \subtitle{- No shallow heating required? -}

%

\author{M. Nava-Callejas\inst{1}\fnmsep\thanks{Corresponding author: martin.javier.nava.callejas@ulb.be}
        \and Cavecchi, Y.\inst{2, 3, 4}
        \and Page, D.\inst{5}
        }

   \institute{{Institut d'Astronomie et d'Astrophysique, Université Libre de Bruxelles, 1050 Bruxelles, Belgique}
   \and {Departamento de Astrof\'isica, Universidad de La Laguna, 38206, San Crist\'obal de La Laguna, Tenerife, Spain}
   \and {Instituto de Astrof\'{i}sica de Canarias, 38205, San Crist\'obal de La Laguna, Tenerife, Spain}
   \and {Center for Nuclear Astrophysics across Messengers (CeNAM), 640 S Shaw Lane, East Lansing, MI 48824, USA}
   \and {Instituto de Astronom\'ia, Universidad Nacional Aut\'onoma de M\'exico, Ciudad de  M\'exico, CDMX 04510, Mexico}
}

   \date{Received September 30, 20XX}

 
  \abstract
   {For the typical modeling of neutron stars cooling after an accretion episode in Low-Mass X-ray Binaries, an extra heating source of unknown physical origin, \textit{the shallow heating}, is invoked in order to account for the inferred high effective temperatures of the star up to hundreds of days after the end of accretion. The amount of the shallow heating generated in the crust is usually taken to be proportional to the accretion rate, although the proportionality constant may change from source to source.}
   {In this paper, we intend to model the effective temperature data of eight outburst episodes from seven different sources (\mxb, \xte, \exo, \igr, \swift, \rxs\ and \ks) without 
   {\it ad hoc}
   shallow heating but accounting for the presence of thermonuclear heating due to the burning of the accreted H/He.\looseness=-9} 
   {We employ the fully relativistic code \nscool, which simulates both the crust and core of a neutron star, equipped with a new boundary condition at the envelope/crust transition which considers thermonuclear heating leakage from the envelope into the crust and depends on the mass accretion rate.}
   {We find that the neutron star cooling for seven out of eight of these outbursts can be well explained with this new boundary condition and without the requirement of 
   {\it ad hoc}
   shallow heating. While the qualitative features of \exo\ cooling curve can be explained, a good fitting still requires additional physics.}
  {}

   \keywords{stars: neutron -- accretion, accretion disks -- X-rays: bursts}

   \maketitle
   \nolinenumbers

\section{Introduction}\label{sec:intro}
When a neutron star (NS) in binary system accretes matter from its companion various processes inject significant amount of heat in its outer layers.
Once the accretion outburst ends, the cooling of the star is visible for hundreds of days \citep{Wijnands_2017}.
In order to explain the early cooling phases, so far it has been necessary to include an additional heating source of unknown origin - dubbed \emph{shallow heating} - during the simulation of the accretion phase \citep[see for instance][]{2009ApJ...698.1020B, Turlione_2015A&A...577A...5T, Potekhin_2025JHEAp..45..116P}, on top of the known processes expected to be at work in the crust \citep{HZ_2008A&A...480..459H, Gusakov2020av}.
In this letter, we argue that the a significant amount of the heat that had been considered as an \emph{ad hoc}
shallow heating is an artifact of how the surface layers are assumed to behave (or, in practice, of simplified surface boundary conditions, BCs, see Appendix \ref{appendix:old_bc} for details) rather than an actual physical mechanism.
Most works modeling mass accretion outbursts treat the neutron star envelope as a boundary condition that is independent of the mass accretion rate $\dot{M}$ and especially of ongoing thermonuclear burning \citep[e.g.][]{2009ApJ...698.1020B, Turlione_2015A&A...577A...5T, Parikh2019}. The latter effects are neglected based on the assumption that  the stellar interior is hotter than the exterior at \emph{all} times. Only a few works have considered the opposite scenario, \citep[\emph{the \twm},][]{1984PASJ...36..199H, 1984ApJ...278..813F, Fujimoto1987op, Miralda-Escude1990wd, Zdunik1992bn}. The implications of the \twm\ for thermal evolution of the neutron star at the end of the accretion period, however, have not been explored yet. The main reason is that full multi-zone simulations of the neutron star evolution during an outburst lasting from months to years are computationally expensive.
To accelerate the calculations of outbursts, in \citet{Nava-Callejas_2025} we introduced a new boundary condition for the envelope of the star which is sensitive to $\dot{M}$ and to the ongoing thermonuclear burning of H/He. This boundary condition was calibrated against time-dependent simulations of the envelope, and is able to account for both the \twm\ and the common paradigm in which heat only flows from the interior into the envelope. In the present work we show that, by using our new boundary condition for simulating outbursts, we can reproduce the cooling behavior of multiple sources without the need of a shallow heating source.
In Section~\ref{sect:methods} we give a brief summary of the numerical code and the boundary conditions we will use for this work. In Section~\ref{sect:results} we present our results and in Section~\ref{sect:disc_concl} we discuss their implications.


\section{Methodology}\label{sect:methods}

To fit the thermal evolution of each NS we employ the latest version of the fully relativistic 1D code \nscool\ \citep{2016ascl.soft09009P}. Further details on the physics included in the code can be found in Appendix~\ref{appendix:nscool_description_2}, while a review of the treatment of the envelope-crust boundary we aim to replace can be found in Appendix~\ref{appendix:old_bc}. Here, we only describe the free parameters of our models and the new scheme we implement for the surface boundary condition.

\nscool\ solves the equations of thermal transport from the center of the star up to a threshold density between the crust and the envelope, typically $\rho_b = 10^{7}$ g cm$^{-3}$. At this threshold density, a boundary condition is imposed. Our ``New BC'' uses the \tblbmdot\ boundary condition described in detail in \citet{Nava-Callejas_2025}, where $T_b$ and $L_b$ are the temperature and flux at the boundary. During the \textit{quiescent} phase, i.e. when $\dot{M} = 0$, this relation reduces to a \tbteff\ relation in which the only relevant parameter is the maximum column depth at which light elements are present, $y_{L}$. Therefore, in this scheme the envelope requires two adjustable parameters: $\dot{M}$ and $y_{L}$. The old boundary condition consisted of only the \tbteff\ relation. We model each outburst as an episode of constant mass accretion rate $\dot{M}$ lasting for a duration of $t_{\mathrm{acc}}$. Once the outburst finishes, we track the evolution of the star for $10^{4}$ days. To avoid numerical instabilities, in our code we smoothly suppress the mass accretion rate for one day before the end of the outburst.

In the present work we fit eight outbursts of seven sources, whose names are displayed in Fig.~\ref{fig:scenario_a_fig1}. We selected these sources since they have different outburst duration times $t_{\mathrm{acc}}$, ranging from $\sim 0.1$ to 24 years, and a similarly ample range of mass accretion rates, between $10^{16}$ to $10^{18}$ g s$^{-1}$. Further details on each source and outburst can be found in Appendix~\ref{appendix:obs_data}. For fitting their cooling curves we use the robust method of Monte Carlo Markov Chain \citep[see for instance][]{Lin_2018, Ootes_2019, Degenaar_2021, Page_2022}. When fitting the cooling curves we use two Scenarios, A and B. In the first we construct cooling curves \emph{without shallow heating}. We thus consider five free parameters: $\dot{M}$, $y_{L}$, the core's temperature \tcore\ and two different impurities, one for the outer and another for the inner crust, $Q_{\mathrm{imp,1}}$ and $Q_{\mathrm{imp,2}}$ respectively, which affect the conductivity. Notice that, although an average mass accretion rate can be inferred from observations, our Scenario A takes $\dot{M}$ as a free parameter. \textit{Its agreement with observations is thus not imposed a priori} and it will work as a sanity check on our results. On the other hand, Scenario B works as a null test: finding out if shallow heating is required despite our well-motivated New BC. Thus, in this scenario we consider eight free parameters: the same five as in Scenario A and three more related to the shallow heating: its strength $Q_{\mathrm{sh}}$, as well as a maximum/minimum density range where it operates, $[\rho_{\mathrm{sh,min}}, \rho_{\mathrm{sh,max}}]$.
Scenario B also takes $\dot{M}$ as a free parameter.


\section{Results}\label{sect:results}

\begin{table*}
\scriptsize
\begin{center}
\begin{tabular}{|p{1.7cm}|p{0.4cm}|p{1.0cm}|p{1.0cm}|p{0.8cm}|p{1.0cm}|p{0.9cm}|p{1.1cm}|p{1.2cm}|p{0.4cm}|p{0.4cm}|p{3.cm}|}
\hline
Source & $t_{\mathrm{acc}}$ & $\langle\dot{M}\rangle^{\dot{M}_{\mathrm{max}}}_{\dot{M}_{\mathrm{min}}}$  & $\dot{M}$ & \tcore & $Q_{\mathrm{imp,1}}$ & $Q_{\mathrm{imp,2}}$ & $\log_{10}y_{L}$ & $\chi^{2}(\mathrm{d.o.f.})$ & AICc & BIC & Refs.\\
& [yr] & $[10^{18}\ \mathrm{g\ s}^{-1}]$ & $[10^{18}\ \mathrm{g\ s}^{-1}]$ & $[10^{7}\ \mathrm{K}]$ & & & $[\mathrm{g\ cm}^{-2}]$ & & & &\\
\hline
%
\mxbi & 2.500  & $0.120^{+0.160}_{-0.050}$ &  $0.21^{+0.11}_{-0.06}$ & $2.07^{+0.34}_{-0.28}$ & $5.23^{+6.65}_{-3.94}$ & $1.32^{+0.82}_{-0.30}$ & $9.06^{+0.85}_{-1.25}$ & & & & \\
& & & 0.16 & 1.95 & 12.10 & 1.01 & 9.94 & 1.96 (1) & $\infty$ & 11 & \cite{Parikh2019}\\
\hline
\mxbii & 1.700  & $0.052^{+0.100}_{-0.005}$ & $0.07^{+0.09}_{-0.04}$ & $1.68^{+1.11}_{-0.63}$ & $7.08^{+25.30}_{-5.83}$ & $2.36^{+3.13}_{-1.26}$ & $7.94^{+1.84}_{-1.77}$ & & & & \\
& & & 0.14 & 1.01 & 12.01 & 2.24 & 6.36 & 0.85 (2) & 71 & 11 & \cite{Parikh2019}\\
\hline
\rxsbrief   & 0.375 & 0.059 & $0.10^{+0.13}_{-0.06}$ & $3.62^{+1.46}_{-1.86}$ & $50.43^{+43.39}_{-48.43}$ & $10.40^{+70.52}_{-9.24}$ & $8.19^{+1.63}_{-1.82}$ & & & & \\
& & & 0.04 & 3.62 & 97.96 & 48.00 & 9.00 & 2.04 (3) & 42 & 12 & \cite{Parikh_2018}\\
\hline
\igrbrief    & 0.170  & $0.19^{+0.20}_{-0.17}$ & $0.84^{+0.80}_{-0.38}$ & $5.58^{+1.97}_{-0.96}$ & $14.86^{+43.23}_{-13.36}$ & $2.52^{+5.81}_{-1.42}$ & $8.34^{+1.51}_{-1.92}$ & & & & \\
& & & 0.55 & 4.87 & 23.40 & 2.72 & 9.61 & 5.39 (4) & 35 & 16 & \cite{Degenaar_2011, Ootes_2019}\\
\hline
\swiftbrief & 0.150  & 0.096 & $0.10^{+0.19}_{-0.07}$ & $6.86^{+2.34}_{-1.51}$ & $8.10^{+67.63}_{-6.85}$ & $8.95^{+72.14}_{-7.70}$ & $8.11^{+1.71}_{-1.89}$ & & & & \\
& & & 0.09 & 8.06 & 72.19 & 28.89 & 7.06 & 0.25 (0) & $\infty$ & 8 & \cite{2015MNRAS.451.2071D}\\
\hline
\ksbrief     & 12.50  & $0.15^{+0.30}_{-0.05}$ & $0.10^{+0.11}_{-0.04}$ & $3.16^{+1.37}_{-0.34}$ & $21.53^{+17.89}_{-18.24}$ & $1.36^{+1.72}_{-0.34}$ & $8.62^{+1.24}_{-2.32}$ & & & & \\
& & & 0.07 & 2.98 & 31.82 & 1.01 & 9.47 & 17.76 (3) & 58 & 28 & \cite{Merritt_2016ApJ...833..186M, Ootes_2016}\\
\hline
\xtebrief    & 1.600  & 1.1 & $0.85^{+0.35}_{-0.28}$ & $12.42^{+3.98}_{-1.92}$ & $20.07^{+74.19}_{-18.67}$ & $2.37^{+8.95}_{-1.30}$ & $8.58^{+1.26}_{-2.29}$ & & & & \\
& & & 0.75 & 13.32 & 99.67 & 1.82 & 8.37 & 14.28 (8) & 33 & 27 & \cite{Parikh_2020}\\
\hline
\exobrief    & 24.00  & 0.030 & $0.05^{+0.001}_{-0.001}$ & $9.60^{+0.03}_{-0.02}$ & $1.23^{+0.81}_{-0.21}$ & $1.01^{+0.04}_{-0.01}$ & $9.99^{+0.01}_{-0.01}$ & & & & \\
& & & 0.05 & 9.60 & 1.01 & 1.00 & 9.97 & 1478.35 (8) & 1497 & 1491 & \cite{Parikh_2020}\\
\hline
\end{tabular}
\caption{\fonto{Results of the MCMC run for Scenario A. For each source we report 
the observed duration of the accretion outburst, $t_\mathrm{acc}$, 
and its average mass accretion rate, $\langle \dot{M}\rangle$. 
We also report the result for the quantities that we fit for our scenario A: 
mass accretion rate, $\dot{M}$, 
red-shifted core temperature, $\tilde{T}_\mathrm{core}$, 
the two impurity factors $Q_\mathrm{imp, 1/2}$ and $\log_{10} y_L$.
For each quantity we report the median values with the ranges from the 5\% and 95\% quantiles (covering the 90\% confidence interval) and the best fitting values on the lower line of each source. 
We list the residuals of the best fitting model with the reduced degrees of freedom in parenthesis and the (modified) Akaike and Bayesian information criteria. 
Finally, for each source we report the references from where the observational data and the average mass accretion rate were taken.
\looseness=-9}}
\label{tab:scenario_a}
\end{center}
\end{table*}

\begin{table*}
\scriptsize
\begin{center}
\begin{tabular}{|p{1.7cm}|p{1.0cm}|p{0.8cm}|p{0.9cm}|p{0.9cm}|p{1.1cm}|p{1.8cm}|p{1.0cm}|p{1.0cm}|p{1.1cm}|p{0.6cm}|p{0.6cm}|}
\hline
Source & $\dot{M}$ & \tcore & $Q_{\mathrm{imp,1}}$ & $Q_{\mathrm{imp,2}}$ & $\log_{10}y_{L}$ & $Q_{\mathrm{sh}}$ & $\log_{10}\rho_{\mathrm{sh, min}}$ & $\log_{10}\rho_{\mathrm{sh,max}}$ & $\chi^{2} (\mathrm{d.o.f.})$ & AICc & BIC\\
& $[10^{18}\ \mathrm{g\ s}^{-1}]$  & $[10^{7}\ \mathrm{K}]$ & & & $[\mathrm{g\ cm}^{-2}]$ & [MeV baryon$^{-1}$] & $[\mathrm{g\ cm}^{-3}]$ & $[\mathrm{g\ cm}^{-3}]$  & & & \\
\hline
%
\mxbi & $0.03^{+0.08}_{-0.01}$ & $2.32^{+0.73}_{-0.48}$ & $6.86^{+10.35}_{-5.59}$ & $2.92^{+3.37}_{-1.81}$ & $8.02^{+1.77}_{-1.70}$ & $3.00^{+2.76}_{-2.49}$ & $8.28^{+1.60}_{-1.08}$ & $9.97^{+1.00}_{-1.95}$ & & & \\
& 0.04 & 2.48 & 6.06 & 2.28 & 7.51 & 2.42 & 9.77 & 10.34 & 0.86 (-) & $\infty$ & 15 \\
\hline
\mxbii & $0.01^{+0.03}_{-0.01}$ & $1.63^{+1.31}_{-0.59}$ & $12.67^{+31.14}_{-11.13}$ & $4.49^{+12.59}_{-3.28}$ & $7.93^{+1.84}_{-1.74}$ & $2.78^{+2.90}_{-2.42}$ & $8.19^{+1.65}_{-1.00}$ & $9.78^{+1.17}_{-1.95}$ & & & \\
& 0.07 & 1.30 & 6.86 & 4.21 & 6.79 & 0.36 & 7.23 & 7.59 & 0.97 (-) & $\infty$ & 17 \\
\hline
\rxsbrief & $0.01^{+0.06}_{-0.01}$ & $3.84^{+1.70}_{-1.05}$ & $10.85^{+70.71}_{-9.51}$ & $12.39^{+71.24}_{-11.12}$ & $8.20^{+1.64}_{-1.92}$ & $2.98^{+2.71}_{-2.64}$ & $8.18^{+1.43}_{-0.99}$ & $9.90^{+1.08}_{-1.77}$ & & & \\
& 0.01 & 3.42 & 3.21 & 1.74 & 9.81 & 2.26 & 7.57 & 10.24 & 1.41 (0) & $\infty$ & 18 \\
\hline
\igrbrief & $0.15^{+0.97}_{-0.13}$ & $5.81^{+2.22}_{-1.64}$ & $18.10^{+68.13}_{-16.83}$ & $7.30^{+58.02}_{-6.04}$ & $8.40^{+1.47}_{-2.04}$ & $0.72^{+4.73}_{-0.70}$ & $8.75^{+1.60}_{-1.52}$ & $10.42^{+0.62}_{-2.42}$ & & & \\
& 0.93 & 6.59 & 88.15 & 4.15 & 6.69 & 0.27 & 10.80 & 10.96 & 4.58 (1) & $\infty$ & 22 \\
\hline
\swiftbrief  & $0.01^{+0.06}_{-0.01}$ & $6.90^{+2.36}_{-1.55}$ & $9.53^{+68.89}_{-8.26}$ & $9.66^{+68.36}_{-8.44}$ & $8.02^{+1.80}_{-1.81}$ & $2.59^{+3.01}_{-2.33}$ & $8.35^{+1.90}_{-1.13}$ & $9.94^{+1.08}_{-1.85}$ & & & \\
& 0.01 & 6.27 & 15.45 & 13.90 & 8.71 & 2.42 & 9.54 & 9.78 & 0.20 (-) & $\infty$ & 13 \\
\hline
\ksbrief & $0.02^{+0.02}_{-0.01}$ & $4.24^{+0.68}_{-1.25}$ & $4.67^{+21.76}_{-3.49}$ & $6.07^{+3.40}_{-4.33}$ & $6.64^{+2.31}_{-0.59}$ & $4.21^{+1.64}_{-2.60}$ & $8.02^{+1.27}_{-0.83}$ & $9.39^{+1.32}_{-1.46}$ & & & \\
& 0.02 & 4.73 & 1.37 & 7.83 & 6.05 & 5.28 & 8.23 & 9.00 & 1.70 (0) & $\infty$ & 18 \\
\hline
\xtebrief & $0.06^{+0.35}_{-0.04}$ & $13.62^{+3.87}_{-2.95}$ & $26.99^{+65.61}_{-25.61}$ & $12.23^{+69.16}_{-10.95}$ & $8.46^{+1.37}_{-2.19}$ & $2.84^{+2.78}_{-2.43}$ & $9.53^{+0.92}_{-1.41}$ & $10.62^{+0.43}_{-0.74}$ & & & \\
& 0.03 & 12.19 & 39.70 & 2.23 & 9.98 & 3.78 & 9.31 & 10.72 & 7.38 (5) & 59 & 28 \\
\hline
\exobrief & $0.01^{+0.001}_{-0.001}$ & $10.07^{+0.10}_{-0.03}$ & $1.21^{+0.69}_{-0.20}$ & $1.04^{+0.14}_{-0.04}$ & $9.71^{+0.06}_{-0.11}$ & $5.78^{+0.21}_{-0.71}$ & $8.28^{+0.41}_{-0.64}$ & $8.82^{+0.46}_{-0.40}$ & & & \\
& 0.01 & 10.06 & 1.00 & 1.00 & 9.75 & 5.93 & 8.35 & 8.66 & 771.84 (5) & 824 & 792 \\
\hline
\end{tabular}
\caption{\fonto{Results of the MCMC run for Scenario B. Same as Table \ref{tab:scenario_a}, but including also the shallow heating, $Q_\mathrm{sh}$,
and the logarithm of the densities between which it is generated, $\log_{10} \rho_\mathrm{sh, min/max}$.}}
\label{tab:scenario_b}
\end{center}
\end{table*}


\begin{figure}
\begin{center}
\includegraphics[width=1.0\linewidth]{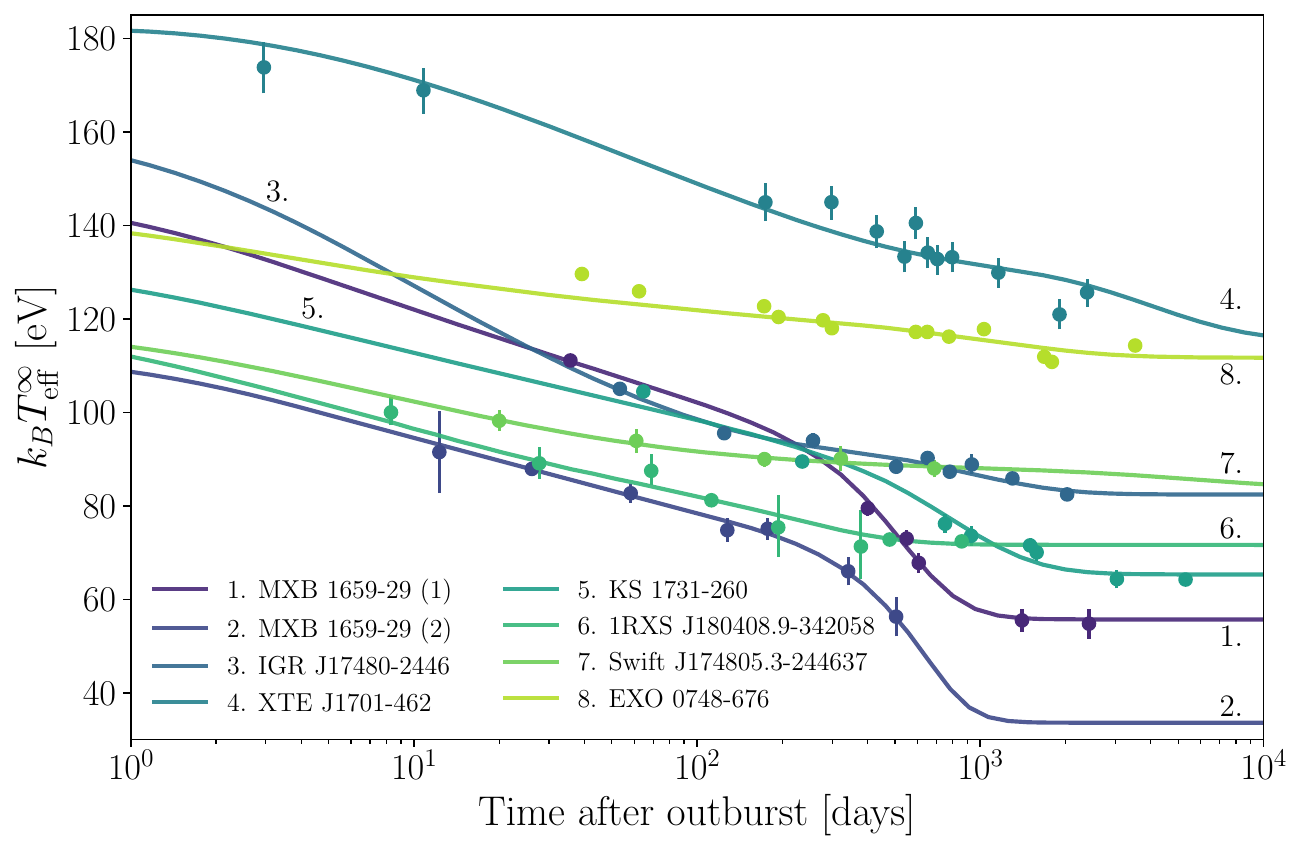}
\end{center}
\caption{\fonto{Best fitting cooling curves for Scenario A. 
}\looseness=-9}
\label{fig:scenario_a_fig1}
\end{figure}


In Table~\ref{tab:scenario_a} we report the best-fit parameters and their 90\% confidence intervals for scenario A, as well as their corresponding $\chi^2$ and the number of degrees of freedom, and
in Fig.~\ref{fig:scenario_a_fig1} we display the best fit curves of Scenario A.
 The extremely good agreement with the data demonstrates that our new BC is sufficient to explain the high $T^{\infty}_{\mathrm{eff}}$ right after the end of the outburst without needing any shallow heating.
In the cases of \rxsbrief, \swiftbrief, \ksbrief, \xtebrief, \exobrief\ and both outbursts from \mxbbrief\ we find a good agreement between the $\dot{M}$ of the fit and the observational average mass accretion rate. Only for \igrbrief\ our model requires a mass accretion rate three to four times larger than what is inferred from observations.
In general, our MCMC simulations favor a decreasing impurity as a function of density: $Q_{\mathrm{imp,1}}$, at the outer crust, is larger than $Q_{\mathrm{imp,2}}$ at the inner crust. The order of magnitude is similar in most cases, $\sim 10$ at the outer crust and $\sim 1$ at the inner one.
Another internal consistency result comes from $y_L$. We see the presence of light elements in most sources is consistent with $y_L \sim 10^8$ g cm$^{-2}$. Since the \tbteff\ relations we use were derived for $^{56}\mathrm{Fe}$ at $y > y_{L}$, our results are consistent with the prediction  that light-elements during explosive or stationary burning of H/He cannot reach densities above $10^8$ g cm$^{-2}$.

The best-fit parameters and their $90\%$ confidence intervals for Scenario B are reported in Table~\ref{tab:scenario_b}. In general, we find mass accretion rates ten times smaller than what is inferred from observations. This decrease in $\dot{M}$, however, is compensated by a high $Q_{\mathrm{sh}}$, which in all cases is of the order of, or in excess of, 5 MeV baryon$^{-1}$. With respect to Scenario A, we find similar values for \tcore, as expected since this parameter is constrained mostly by the cooling behavior at $t\geq 10^{2}$ days. While the impurities are different between the two scenarios, we also note that in some cases of Scenario B the required impurity in the inner crust exceeds that for the outer one, as is the case for \swiftbrief\ and \ksbrief.\looseness=-9 The corner plots for both Scenarios can be found in Appendix~\ref{appendix:corners}.


\section{Discussion and Conclusions}\label{sect:disc_concl}

In this letter we explored the possibility of removing the shallow heating from the simulations of neutron star cooling after an outburst episode. We instead employed a new envelope boundary condition which does account for the effects of thermonuclear burning of H/He when $\dot{M}\neq 0$, allowing for energy to flow from the envelope into the crust - the \twm\ \citep{1984PASJ...36..199H,Fujimoto1987op,Nava-Callejas_2025}. 

In Scenario A, \emph{without shallow heating}, we fitted our sample of observations employing five free parameters, among them $\dot{M}$. In Scenario B, we tested the possibility that even with our New BC, shallow heating might still be required. 
We find that Scenario A is capable of fitting the observational data for the majority of the sources. Furthermore, the predicted mass accretion rate from our MCMC simulations, which was set as a free parameter, reproduces very well the average mass accretion rate inferred from observations. 

When an F-test is possible, comparing the two scenarios always prefers Scenario A. In both Table~\ref{tab:scenario_a} and~\ref{tab:scenario_b} we report the values of the Akaike and Bayesian Information Criteria \citep{Feigelson_2012} for each outburst. These can be used to quantify whether the extra free parameters of Scenario B are justified by a substantial reduction of the residuals in the $\chi^2$. A model with a lower value is preferred. When comparing the AICc and BIC values, Scenario A is favored for 5 out of 7 sources - \mxbbrief, \rxsbrief, \igrbrief{} and \xtebrief (for this the BIC is inconclusive, but the AICc favors the first scenario). For \ksbrief{} each test favors a different scenario, while for \exobrief\ both criteria agree on Scenario B being the most favorable one.
However, the values required by the fits of Scenario B are unphysical: $\dot{M}$ must be one order of magnitude smaller than what is reported by observations and the strength of the shallow heating is of order 4 MeV baryon$^{-1}$.\looseness=-9

While both outbursts from \mxbbrief\ were treated as independent, \tcore, $Q_{\mathrm{1,2}}$ and $\log_{10}y_{L}$ are consistent between outbursts at the $90\%$ confidence level. This is in agreement with Scenario C of \citet{Parikh2019} who linked the parameters for both fits, save for $\dot{M}$ whose average value is different in each outbursts. 
\rxsbrief\ is a particularly interesting source: in addition to the good agreement between our $\dot{M}$ and the observational average, previous studies \citep[e.g.][]{Parikh_2017, Parikh_2018} had pointed out that the shallow heating mechanism has a more prominent role than the deep crustal heating in explaining the evolution of the star during and after the outburst. Our model is capable of explaining both phases naturally: since the short-lived outburst duration is comparable to the heat diffusion timescale of the outer crust ($\geq 100$ days, e.g. \citealt{2009ApJ...698.1020B}) heat flowing from the envelope has ample time to propagate into the crust during the accretion phase. Thus, at the end of the outburst the crust remains sufficiently hot as to lead to an equally high $T^{\infty}_\mathrm{eff}$.
In the case of \swiftbrief, the source with the shortest outburst, previous modelings have explored configurations with and without shallow heating, albeit with the old envelope boundary condition \citep[e.g.][]{Bahramian_2014, 2015MNRAS.451.2071D}. While both approaches can reproduce the post-outburst cooling, the inclusion of shallow heating slightly improves the agreement with the inferred pre-outburst $k_{B}T^{\infty}_\mathrm{eff} = 89.7\pm 1.7$ eV. In our Scenario A, both observational constraints are reproduced without invoking shallow heating since we obtain a good fit and a pre-outburst $k_{B}T^{\infty}$ value of 90 eV.
In the case of \ks, we reach a similar conclusion as in \citet{2013PhRvL.111x1102P} and \citet{Jain_2025} favoring a decreasing $Q_{\mathrm{imp}}$ as a function of density. This trend, however, is observed in all fits of Scenario A. While our best fit $\dot{M}$ and the observationally inferred average $\langle \dot{M}\rangle$ already agree well, the actual outburst from \ks\ shows large variability (see Appendix~\ref{appendix:obs_data}). Therefore, a variable $\dot{M}$ may lead to an improved fit, in particular considering that a different $\dot{M}$ provides a different amount of injected heating at the envelope-crust boundary (Fig. 2 from~\citealt{Nava-Callejas_2025}).
\xtebrief\ and \exobrief\ are interesting cases since both have exhibited a late-time increase in $T^{\infty}_{\mathrm{eff}}$, which in principle can be due to a physical mechanism different from residual accretion \citep{Parikh_2020}. Scenario A is able to reproduce well the cooling curve \xtebrief, while it remains unsatisfactory in the case of \exobrief. One caveat to consider is that also \exobrief, similarly to \ksbrief, has been reported to be a neutron star of $2M_{\odot}$ \citep{Ozel_2006, Ozel_2012, Knight_2022}, hence further effects might be at play in both \exobrief\ as in \ksbrief; in particular taking into account that both sources have $t_{\mathrm{acc}}\geq 10$ yr.
While Scenario B does a better job at fitting the cooling curve of \exobrief, the shallow heating strength is unusually large. The values of $Q_\mathrm{imp}$, however, are surprisingly similar in both Scenarios, \exobrief\ being the source with the smallest impurity.

All these results indicate that shallow heating may not be needed to explain the cooling behaviour of NSs in transient sources.
A clue in this direction was posed in \cite{Page_2022}, in which the best fitting of multiple outburst from the transient source MAXI J0556-332 was consistent with a single shallow heating located at the envelope/crust boundary. Since the envelope in this work was treated using the \tbteff{} scheme, such result showed that accounting for extra processes in the deep outer layers of that NS could reduce the need for an exceptionally high shallow heating. Similarly, \citet{cave2025} showed that the stabilization of type I bursts (explosive thermonuclear runaways on the surface of accreting NSs in low mass X-ray binaries) at mass accretion rates lower than theoretically predicted, previously explained through some high energy flux into the burning ocean from below, can in fact be explained simply accounting for changes in the disc structure as a function of global mass accretion rate, that affect the local composition and accretion rate driving the bursts.

While the agreement between our models without shallow heating and the observational data is good, there are several aspects that could be improved.
A first step may be including time-dependent mass accretion rate, modeled after the observations of each source, in particular in the cases of \ksbrief\ or\ \igrbrief, where significant variability is observed throughout their outburst. Second, allowing for different values of neutron star mass and radius, fixed here to the canonical values of $1.4 M_{\odot}$ and 11.56 km, will account in a more consistent way additional physical processes in the interior of the more massive stars. Third, one may model $Q_\mathrm{imp}$ as a continuous function of depth, following the results from  simulations that push the ashes from rp-process or superbursts at high depths \citep[e.g.][]{2013PhRvL.111x1102P, Potekhin_2025JHEAp..45..116P, Jain_2025}. The fact that our fits already qualitatively reproduce such trends is a further indication of their validity.
Nonetheless, despite these possible improvements, our results already strongly suggest that thermonuclear heating in the envelope of NSs plays a bigger role in shaping their cooling curves than previously believed, to the point that shallow heating may not be needed for many, if not all sources.


\begin{acknowledgements}
M.N.-C. acknowledges support by the Fonds de la Recherche Scientifique-FNRS under Grant No IISN 4.4502.19. Y.C. acknowledges support from the grant RYC2021-032718-I, financed by MCIN/AEI/10.13039/501100011033 and the European Union NextGenerationEU/PRTR.
D.P.'s work is supported by an UNAM-DGAPA grant PAPIIT-IN114424. 
\end{acknowledgements}

\vspace{-0.8cm}

\bibliographystyle{aa}
\bibliography{nscool_nosh}

\begin{appendix}

\section{On the origins of the shallow heating}\label{appendix:old_bc}

For simulating the evolution of a NS temperature during and after transient episodes of mass accretion \emph{with shallow heating}, most authors \citep[e.g.][]{2009ApJ...698.1020B, Turlione_2015A&A...577A...5T, Ootes_2016, Parikh2019} implicitly assume that \emph{during outburst and quiescence, the envelope of the neutron star does not contain internal heating or cooling sources}. Consequently, the luminosity is constant across the envelope and this region of the star can be replaced by a \tbteff\ relation \citep{1982ApJ...259L..19G, Potekhin:1997mn}\footnote{Note that \cite{Potekhin_2021A&A...645A.102P} despite not using an explicit \tbteff\ relation, still find $L_b >0$ given they keep the assumption of an envelope without heating sources.}. Therefore, $T_b$ is only a function of the effective temperature at the surface, $T_{\mathrm{eff}}$, and so is the luminosity at the envelope-crust boundary, $L_b = 4\pi R^{2}T^{4}_{\mathrm{eff}} >0$.

In general, $T_{b}\propto T^{4}_{\mathrm{eff}}$. As a corollary, if by the end of the outburst we require the star to have a high $T_{\mathrm{eff}}$, e.g. $\sim 10^{6}$ K, then the corresponding $T_b$ must be high as well, e.g. $T_b \geq 3-6\times 10^{8}$ K. However, the afore mentioned assumption leads to the requirement of a heating source capable of producing such temperature (and luminosity).
While these \tbteff\ relations are sensitive to the chemical composition of the envelope, they are insensitive to both the mass accretion rate $\dot{M}$ and the details of thermonuclear burning. Such neglect results in a physically inconsistent model, especially under conditions when the burning of  H, He and C is a relevant source of heat at densities lower than $10^{8}$ g cm$^{-3}$.

That the shallow heating may have been a boundary condition artifact could be spotted already considering the ways in which it has been implemented in the literature. There are two approaches:
an \emph{implicit} shallow heating, which results from keeping the envelope-crust boundary temperature $T_b$ fixed during the whole outburst \citep[as in, e.g.,][]{2009ApJ...698.1020B, Turlione_2015A&A...577A...5T}, or an \emph{explicit} shallow heating, produced by distributing a fixed amount of energy along a density (or pressure or radial) interval
\citep[see for instance][]{Ootes_2016, Parikh2019}. In both cases, $T_b$ needed to be sufficiently high by the end of the outburst due to the assumption mentioned above, which neglected heat sinks or sources.


\section{On the physics of \nscool}\label{appendix:nscool_description_2}

For the neutron star core we consider the APR-EOS \citep{Akmal:1998vu}, consistent with having a maximum mass of $2M_{\odot}$ as well as with values of tidal deformability derived from gravitational waves constraints \citep{Abbott_2017_GW}. For the crust of the neutron star, as well as for the deep crustal heating (present by construction only when $\dot{M}\neq 0$) we consider the accreted-EOS from \cite{HZ_2008A&A...480..459H}, consisting of layers of material synthesized via chains of non-equilibrium electron captures over a fiducial $^{56}$Fe layer at $\sim 10^{8}$ g cm$^{-3}$. For the neutron $^{1}\mathrm{S}_{0}$ superfluidity we use the SFB gap \citep{SCHWENK2003191}, while for proton $^{1}\mathrm{S}_{0}$ superconductivity we employ the gap from CCDK \citep{Chen:1993wu}. Typically, \nscool\ assumes a \tbteff\ relation with chemical composition of $^{4}$He, $^{12}$C and $^{56}$Fe, and takes $y_{L} = y_{^{12}\text{C}}$, fixing $y_{^{4}\text{He}} =  y_{^{12}\text{C}}/10$ according to the ratios found in \cite{2024RASTI...3..800N}. In all of our models, the mass and radius of the neutron star are fixed to the APR Core-EOS values of $1.4 M_{\odot}$ and $11.56$ km. Consequently, we also fix the neutron drip point, onset of pasta phase and core-inner crust densities. They take, respectively, the following values: $6.2\times 10^{11}$ g cm$^{-3}$, $8\times 10^{13}$ g cm$^{-3}$ and $1.5\times 10^{14}$ g cm$^{-3}$.
The conductivity of the crust is mainly controlled by the electron-ion scattering process, which is sensitive to the chemical composition. Self-consistent many-body calculations of this conductivity are complicated due to the strong coupling of nuclides \citep{Roggero_2016}. In order to simplify these calculations, most works assume the crust is mainly composed of a single-nuclear species inside which there are multiple but less-abundant species, the so called impurities. As a result, the conductivity can be controlled by a single parameter $Q_\mathrm{imp}$ \citep[see][and references therein]{2013PhRvL.111x1102P, Nava-Callejas_2025}. Since the crust composition depends on density as well, the impurity can vary from layer to layer. In this work we consider only two impurities, one for the inner and another for the outer crust.
Regarding the deep crustal heating, to date there exist two competing scenarios. The one used in this work comes from \cite{HZ_2008A&A...480..459H} and provides up to a few MeV per baryon at different layers in the inner and outer crust, in rather good agreement with large networks of reactions \citep[as, e.g. those of][]{Gupta_2007ApJ...662.1188G}. The second scenario, from \citet{Gusakov2020av}, includes neutron diffusion in the calculations and the amount of deep crustal heating decreases up to one order of magnitude with respect to \citet{HZ_2008A&A...480..459H}. Note, however, that the exact choice of the deep crustal heating is not very important, since both scenarios have been insufficient to reproduce the observations \citep[e.g.][]{Ootes_2019A&A...630A..95O} without shallow heating. Furthermore, in sources such as \rxs\ the complete removal of deep crustal heating has shown negligible impact in the modeling of the evolution as long as there is a shallow heating, i.e. the very mechanism we aim to replace with our New BC. Therefore, we argue that reducing the amount of deep crustal heating should have a minor effect in our conclusions regarding the usage of a new boundary condition. Nevertheless, we defer to an upcoming work the examination of the impact of changing the amount of deep crustal heating over the models with our new boundary condition.


\section{Observational data}\label{appendix:obs_data}

\subsection{\mxb}\label{subsubsect:mxb}

To date, this source has exhibited three outbursts, although only for the last two - typically labeled as OI and OII - there were monitoring campaigns of the cooling phase \citep{Wijnands_2002, Turlione_2015A&A...577A...5T, Parikh2019}.
The data we use for both OI and OII come from \cite{Parikh2019}. These authors report the fluence for both outbursts and display the bolometric flux, from where we compute the average mass accretion rate for each outburst: $1.2 \times 10^{17}$ g s$^{-1}$ and $5.2 \times 10^{16}$ g s$^{-1}$. The OI lightcurve shows a rather constant behavior, while OII displays more variability during the first months of the outburst. We did not find mass and radius constraints inthe literature, although it has been speculated that \mxb\ might host a neutron star massive enough as to sustain Urca processes \citep{Ozel_2006, Brown_2018}.


\subsection{\ks}\label{subsubsect:ks}

During its 12.5 years of mass accretion, the outburst lightcurve of this source displays up to four regimes \citep[see Fig.1 from][]{Ootes_2016}: from MJD 47500 to 48500, $\dot{M}\sim 1.2 \cdot 10^{17}$ g s$^{-1}$; from MJD 48500 to 50000, $\dot{M}\sim 2.3\cdot 10^{17}$ g s$^{-1}$; from 50000 to 51000, $\dot{M}\sim 3.5 \cdot 10^{17}$ g s$^{-1}$ and, finally, from 51000 to 52000 $\dot{M}\sim 1.2 \cdot 10^{17}$ g s$^{-1}$ again. Considering the whole observational data, $\langle\dot{M}\rangle = 1 - 1.5 \cdot 10^{17}$ g s$^{-1}$  \citep{Galloway_2008, Ootes_2016}.
For the cooling simulation we consider the latest reported dataset from \cite{Merritt_2016ApJ...833..186M}, while the mass accretion rate range inferred from observations comes from \cite{Ootes_2016}. 
According to \citet{Ozel_2012}, \ksbrief\ is compatible with being a $2\, M_{\odot}$ neutron star. However, in our model we treated it as a $1.4\, M_{\odot}$ star in order to be compatible with the \tblbmdot\ boundary condition we employed, which is calculated for such a mass. Consequently, effects such as Direct Urca or $^{3}\mathrm{P}_{2}$ pairing might play a role in this source which we did not account for.


\subsection{\xte}\label{subsubsect:xte}
Discovered in 2006, this source exhibited a short outburst of 1.6 yr at a very high mass accretion rate, close to the Eddington limit \citep[e.g.][and references therein]{2011ApJ...736..162F}. The outburst lightcurve exhibits a fast rise and a subsequent exponential decay over 100 days. It then remained at an almost constant mass accretion rate for 400 days, and eventually the mass accretion rate exponentially decayed in the last 25 days. We take $\langle\dot{M}\rangle = 1.1 \cdot 10^{18}$ g s$^{-1}$, $t_{\mathrm{acc}} = 1.6$ yr, and the cooling data from \cite{Parikh_2020}.

\subsection{\rxs}\label{subsubsect:1rxs}
Since its discovery in 1990, this source has shown multiple outbursts, out of which we model the latest one, which took place in 2015 \citep[see][and references therein]{Degenaar_2016_rxs, Parikh_2017}.
The outburst lightcurve of \rxsbrief\ showed an almost constant mass accretion rate for the first 70 days, after which it displays an increase in the intensity - transitioning from the hard to the soft state - followed by a quick drop in the subsequent 60 days until the source returned to quiescence \citep{Degenaar_2016_rxs, Parikh_2017}.
For this source we take $t_{\mathrm{acc}} = 4.5$ months and $\langle\dot{M}\rangle = 5.9\cdot 10^{16}$ g s$^{-1}$, as reported in \cite{Parikh_2017}. The observational data is taken from  \cite{Parikh_2018}. 

\subsection{\igr}\label{subsubsect:igr}
This is the second low mass X-ray binary transient source detected in the globular cluster of Terzan-5. Observational constraints indicate a source with a high magnetic field.
Its outburst lightcurve displays a mass accretion build up during the first ten days, followed by a decay until it almost stabilizes at a constant value. Along these variations, several X-ray bursts and mHz QPOs were detected \citep{Degenaar_2011, Linares_2012, Degenaar_2013}.
It has been suggested that the cooling of \igrbrief{} is compatible with the neutron star having a very impure inner crust, i.e. having a reduced conductivity \citep{Turlione_2015A&A...577A...5T, Ootes_2019}. While the presence of a buried magnetic field has been proposed as an explanation, a self-consistent 2D cooling simulation is still missing.

The observational data for this source is taken from \cite{Ootes_2019}. We adopt $t_\mathrm{acc} = 0.17$ yr and fix $\langle\dot{M}\rangle = 2 \cdot 10^{17}$ g s$^{-1}$ (compatible with the estimate of $1.65\cdot10^{17}$ g s$^{-1}$ from \cite{Ootes_2019}, $1.9 \cdot 10^{17}$ g s$^{-1}$ from \cite{Degenaar_2013} and $2 \cdot 10^{17}$ g s$^{-1}$ from \cite{Turlione_2015A&A...577A...5T}).

\subsection{\swift}\label{subsubsect:swift}
This is the third low mass X-ray binary transient source detected in the globular cluster of Terzan-5 \citep{Bahramian_2014, 2015MNRAS.451.2071D}. Discovered in early July 2012, this source remained active for $\sim 8$ weeks, i.e. $\sim 0.15$ yr. Due to the absence of variations in the thermal spectrum during the pre-outburst phase, the effective temperature at infinity of the neutron star before the outburst was constrained to be $89.7\pm 1.7$ eV. On the other hand, in the post-oubturst period five observations are available, taken between September 2012 and July 2014 with Chandra \citep{2015MNRAS.451.2071D}. In this later work the crust cooling modeling of \swiftbrief\ was compatible with the possibility of not requiring shallow heating at all, although the late time cooling behavior was better explained by having a cold core and including a shallow heating of up to $0.85$ MeV per baryon.
Following \cite{2015MNRAS.451.2071D} we take $t_{\mathrm{acc}}$ = 0.15 yr and $\langle\dot{M}\rangle$ $=9.6\cdot 10^{16}$ g s$^{-1}$. As such, it is the outburst with the shortest duration among those modelled in this work.

\subsection{\exo}\label{subsubsect:exo}
This is the transient with the largest outburst among the sources studied in this work, $t_\mathrm{acc} = 24$ years. Coincidentally, it also has the lowest average mass accretion rate $\langle\dot{M}\rangle \sim 3\cdot10^{16}$ g s$^{-1}$, since across all observations its mass accretion rate remained consistently low. Multiple X-ray bursts have been observed and based on their properties \citet{Ozel_2006} concluded that the neutron star in this LMXB is very massive, $\sim 2\, M_{\odot}$.
For this source, we consider the most recent analysis of the observational data by \citet{Parikh_2020}. In this work the error bars were considerably reduced with respect to previous work, as displayed in, e.g., \citet{Turlione_2015A&A...577A...5T}.


\section{Corner plots}\label{appendix:corners}

\begin{figure*}
\begin{center}
\includegraphics[width=1.0\linewidth]{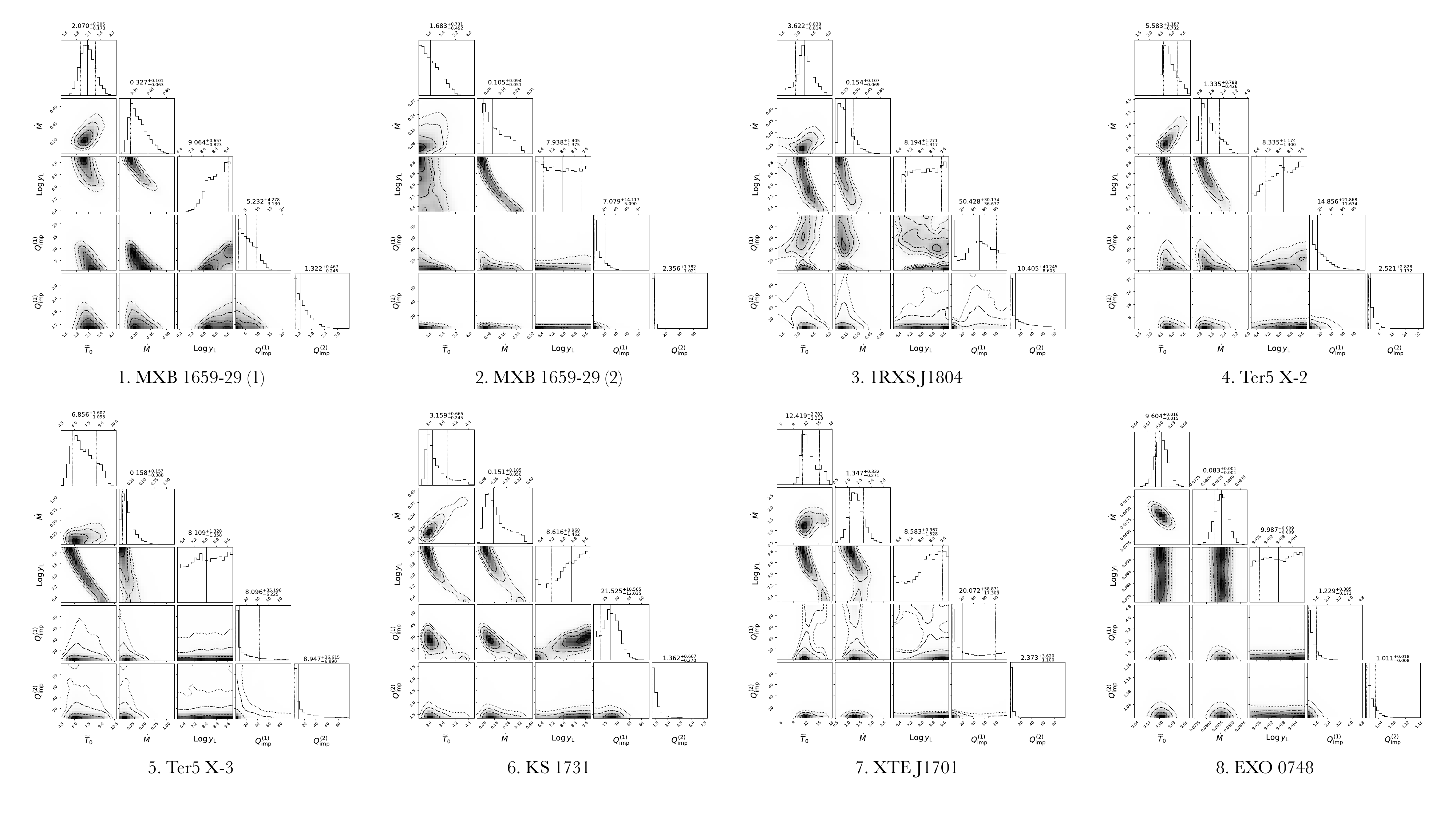}
\end{center}
\caption{\fonto{Corner plots for Scenario A.}\looseness=-9}
\label{fig:scenario_a_corners}
\end{figure*}

\begin{figure*}
\begin{center}
\includegraphics[width=1.0\linewidth]{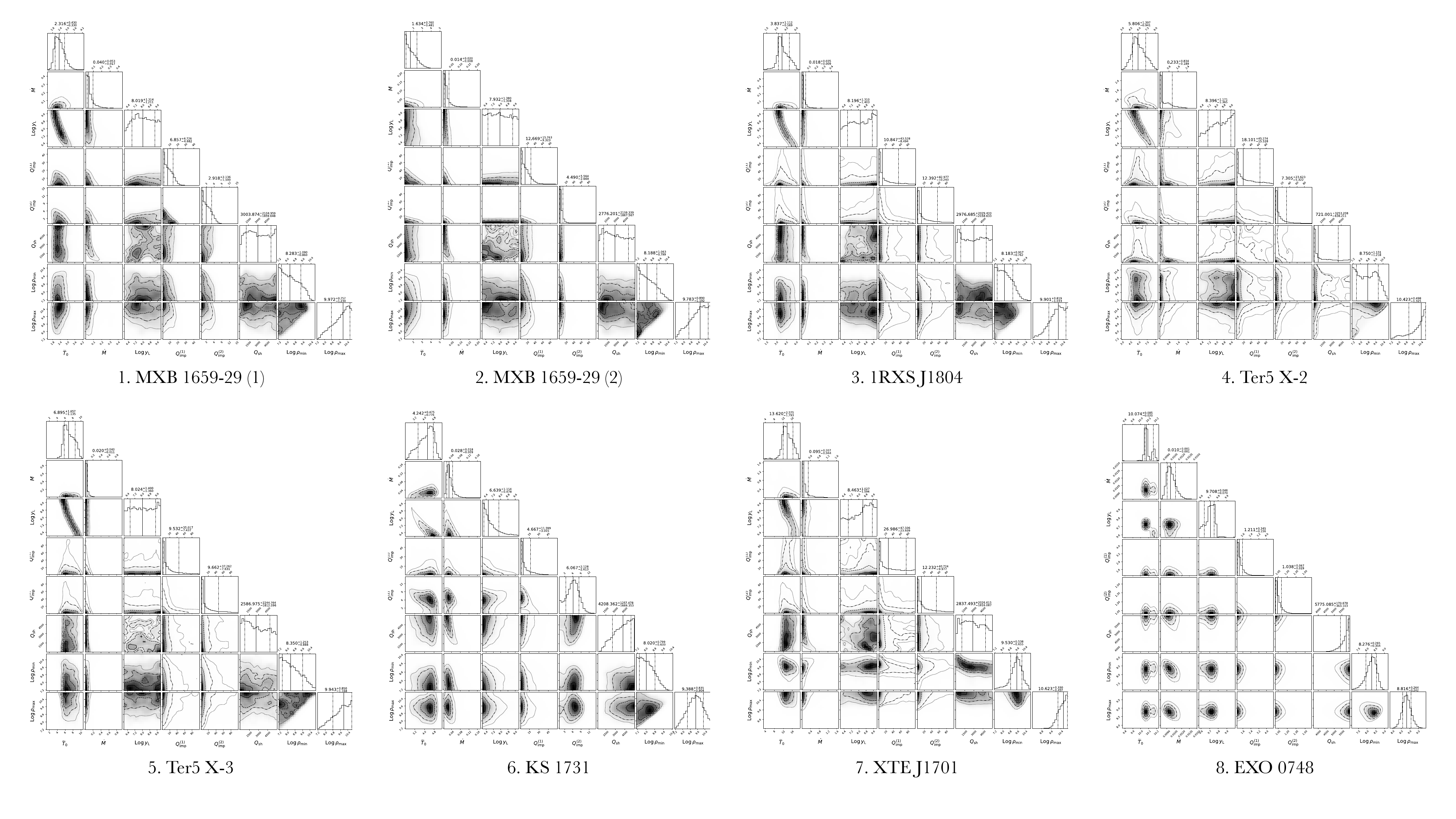}
\end{center}
\caption{\fonto{Corner plots for Scenario B.}\looseness=-9}
\label{fig:scenario_b_corners}
\end{figure*}

In this appendix we show the corner plots resulting from the fits of each source for Scenario A (Fig. \ref{fig:scenario_a_corners}) and Scenario B (Fig. \ref{fig:scenario_b_corners}). The ranges reported in Tables \ref{tab:scenario_a} and \ref{tab:scenario_b} are based on the posteriors shown in these figures. The chains in the MCMC have all converged (they are longer than 50 times the autocorrelation length for each parameter). One can appreciate that some parameters, such as $y_L$, are sometimes unimportant for the fit, having flat posteriors.

\end{appendix}

\end{document}